\documentclass[11pt]{article}
\usepackage{epsfig}

\begin{document}
\def\beq{\begin{equation}}
\def\eeq{\end{equation}}
\vskip30pt

\begin{center}

\rightline{April 2001}
\rightline{UCOFIS 2/01}
\vskip 0.5cm
{\large\bf PARTON DENSITIES AND DIPOLE CROSS-SECTIONS AT SMALL $X$ IN
LARGE NUCLEI }
\vskip20pt
{N. Armesto\\
{\it Departamento de F\'{\i}sica, M\'odulo C2, Planta baja, Campus de
Rabanales,}\\ {\it Universidad de C\'ordoba, E-14071 C\'ordoba, Spain}\\\vskip
0.2cm
and\\ \vskip 0.2cm
M. A. Braun
\\ {\it Department of High-Energy Physics, St. Petersburg
University,}\\
{\it 198504 St. Petersburg, Russia}}
\end{center}

\vskip30pt
\begin{abstract}
Unintegrated gluon densities in nuclei, dipole-nucleus cross-sections and quark
densities are numerically investigated in the high-colour limit, with
the scattering on a heavy nucleus exactly described by the sum of fan diagrams
of BFKL pomerons. The initial condition for the evolution in rapidity
is quickly forgotten, and the gluon density presents a "supersaturation"
pattern, as previous studies indicated. Both dipole-nucleus cross-sections and
quark densities present the expected saturation features. Identifying the
position in transverse momentum $l$ of the maximum of the gluon distribution
with the saturation momentum $Q_s(y,b)$,
at large rapidities all distributions
depend only on one variable $[l/Q_s(y,b)]$ or $[rQ_s(y,b)]$.
\end{abstract}

\section{Introduction}

In view of the current and forthcoming experimental investigations of the
strong interaction with heavy nuclei at high energies, much attention
has lately been devoted to the theoretical study of parton distributions
inside a heavy nucleus at small values of the scaling variable $x$.
It turned out that a particularly transparent approach follows from the
colour dipole picture [1,2], in which the interaction of a probe with
a target is presented via the interaction of the latter with a colour
dipole, convoluted with the distribution of colour dipoles in the probe.
In such a picture the fundamental quantity is the  cross-section
$\sigma (Y,r)$
for the interaction of a colour dipole of a given transverse radius $r$
with a target at a given rapidity $Y$. Much popularity has obtained the
idea of "saturation", which in terms of the cross-section $\sigma$
implies that at high energies the cross-section $\sigma(Y,r)$
tends to a constant independent of $r$ [3]. A particular ansatz
chosen in [3] for the scattering of a dipole on the proton,
\beq
\sigma_p(Y,r)=\sigma_0\left(1-e^{-\hat{r}^2}\right),\ \
\hat{r}=\frac{r}{2R_0(Y)}\ ,
\eeq
with $\sigma_0\simeq 23$ mb and $R_0$ diminishing with increasing
rapidity $Y$,
leads to a good description of the DIS data for the
proton below $x=0.01$.

A less phenomenological treatment can be applied for the heavy nucleus
target. In the framework of the colour dipole model
it follows that, in the high-colour limit $N_c\to\infty$, the
scattering on a heavy nucleus is exactly described by the sum of fan
diagrams constructed of BFKL pomerons, each of them splitting into two.
The resulting equation for the colour dipole cross-section on the
nucleus [4-7] was numerically solved in [7]. The gluon density introduced in
[7] revealed a "supersaturation" behaviour, tending to zero at any fixed
momentum $k$ as  $Y\to\infty$. As a function of $\ln k$ it proved
to have a form of a soliton wave moving to the right with a constant
velocity as $Y$ increases. A more ambitious project is currently
developed by the McLerran group, which can, in principle, lead to
a description which does not employ the large number of colors limit
[8-10].
Admitting that such an improvement is highly desirable, we
think that the $N_c\to\infty$ approach is much more feasible and
can give a clear hint on the qualitative behaviour of
parton densities and cross-sections for the heavy nucleus target.

In this paper we continue studying the numerical solution of the
BFKL fan diagram equation started in [7], with more precision
and more attention to  rapidities available at present or in the near future.
We compare  solutions obtained from a purely theoretical initial function (as
in [7]) and from a phenomenologically supported one (like Eq. (1)). We find that
the initial form is very quickly forgotten by the equation, so that
at rapidities of the order 10 the solution becomes independent
of the chosen initial form. The behaviour of the solution at
large energies is completely determined by the scale $Q_s(Y,b)$,
depending on the energy and impact parameter $b$, at which the gluon density
reaches its maximum value, so that both
the parton densities and the dipole cross-section become
universal functions of momentum or coordinate scaled with $Q_s$.
One may consider $Q_s$ as a "saturation momentum" introduced in
[4,8,11]. The $Y$ and $b$ dependence of $Q$ which follow from our
numerical studies can be fitted by a simple formula:
\beq
\ln Q_s=a+cy+d\ln[AT(b)],\ \ \
y=\frac{N_c\alpha_s}{\pi}Y,
\eeq
with
\[ c=2.06,\ \ d=0.62 \pm 0.07,\]
$Q_s$ in GeV/c, and $T(b)$ the nuclear profile function in
(GeV/c)$^{-2}$ defined so that $\int d^2b\, T(b)=1$;
thus, $AT(b) \propto A^{1/3}$.
Note that the value of $c$ results lower than in our previous run
devoted mostly to asymptotic energies and agrees with the predictions in [12]
based on asymptotic estimates.
The somewhat unexpected value of $d$ seems to slightly depend  on the choice of
the initial function (hence the $\pm 0.07$). It implies that $Q_s\propto
A^\alpha$, $\alpha\simeq 2/9$.  The dipole cross-section on the nucleus
at fixed impact
parameter
exhibits the
expected behaviour, tending to unity at each $r$ as $Y\to\infty$.
The quark density follows the pattern of saturation,
tending to a constant value at small momenta, in full agreement
with the predictions of [13].

\section{The evolution equation and initial conditions}

As mentioned, in the colour dipole approach the cross-section
of a probe ($P)$ on the nucleus ($A$)
is presented via the dipole cross-section:
\beq
\sigma_{PA}(Y)=\int d^2r\rho_P(r)\sigma_A(Y,r).
\eeq
In its turn, the dipole cross-section on the nucleus is an integral over
the impact parameter:
\beq
\sigma_A(Y,r)=2\int d^2b\Phi(Y,r,b),
\eeq
where evidently $2\Phi$ has a meaning of a cross-section at fixed
impact parameter. The evolution equation in $Y$ can be most conveniently
written for the function
\beq
\phi(Y,r,b)=\frac{1}{2\pi r^2}\Phi(Y,r,b)
\eeq
in momentum space, where it reads [7]
\beq
\left(\frac{\partial}{\partial y}+H_{BFKL}\right)\phi(y,q,b)=
-\phi^2(y,q,b),
\eeq
with a rescaled rapidity $y$ given in Eq. (2) and $H_{BFKL}$ the BFKL
Hamiltonian.

Special attention has to be devoted to the initial function
$\phi(y_0,q,b)=\phi_0$ at $y=y_0$, from which value one is starting the
evolution. In the framework of the pure BFKL approach, with $N_c\to\infty$
and $A$ fixed, one should choose $y_0=0$
(or any finite $y$) and take for $\phi_0$ the contribution of the
pure two gluon exchange with a single nucleon inside the nucleus.
If $A$ is large ($A^{1/3}$ of the same order as $N_c$) then one should
add all multiple interactions inside the nucleus, which sum into a
Glauber cross-section:
\beq
\Phi_0(r,b)=1-e^{-AT(b)\sigma_p(0,r)}.
\eeq
Here $\sigma_p(0,r)$ is the dipole-proton cross-section generated
by the two-gluon exchange:
\beq
\sigma_p(0,r)=\frac{1}{2}g^4\int d^2r^\prime G(0,r,r^\prime)\rho_p(r^\prime),
\eeq
where $\rho_p(r)$ is the colour dipole density in the proton and
$G(0,r,r^\prime)$ is the BFKL Green function at $Y=0$:
\beq
G(0,r,r^\prime)=\frac{rr^\prime}
{8\pi}\frac{r_<}{r_>}\left(1+\ln\frac{r_>}{r_<}\right),\ \ r_{>(<)}={\rm
max(min)} \{r,r^\prime\}.
\eeq
Obviously the density $\rho_p$ is non-perturbative and not known.
As in [7], to simplify the calculations we choose $\rho_p$ to be
a normalized Yukawa distribution
\beq
\rho_p(r)=\frac{\mu}{2\pi}\frac{e^{-\mu r}}{r}\ ,
\eeq
with $\mu=0.3$ GeV adjusted to the nucleon radius value.
With Eq. (10) we find
\beq
AT(b)\sigma_p(0,r)=B\left[2{\rm C}-1+2\ln \tilde{r}-{\rm
Ei}(-\tilde{r})\left(2+\tilde{r}^2\right)+e^{-\tilde{r}}(1-\tilde{r})\right],
\ \
\tilde{r}=\mu r,
\eeq
where C is the Euler constant and
dimensionless $B$ carries all the information about the nucleus.
For Pb at the center ($b=0$) it reaches the value 0.12.
Of course,
this choice of $\rho_p$ may look rather arbitrary. We noted in [7]
that calculations demonstrated a certain indifference of the evolution
equation to the choice of the initial function at high enough $y$.
To see this more clearly, in this run we also used an alternative initial
function, more adjusted to the existing experimental data at comparatively
small rapidities. A natural choice would be to take directly (1). However
its analytic form makes it rather difficult to pass to the momentum
space employed in our method of evolution. Therefore we choose a
slightly different form for the phenomenologically motivated
initial dipole cross-section on the nucleon:
\beq
AT(b)\sigma^{ph}_p=B
\left(1-e^{-\hat{r}}\right)^2,
\eeq
with the same parameters as in [3] and $Y_0$ corresponding to $x=0.01$.
The cross-section in Eq. (12) has the same asymptotic behaviour as
that in Eq. (1), both at
$r\to 0$ and
$r\to\infty$. For finite $r$ it is slightly different, but
this difference is not too large and, after all, the choice (1) is
also quite {\it ad hoc}. Note that the asymptotic behaviour of the
phenomenological cross-section is different from that of
Eq. (11), which both at $r\to 0$ and
$r\to\infty$ contains an extra $|\ln r|$ factor. The dimensionless
$B$ in Eq. (12) results substantially larger than for Eq. (11) and is of the
order of 5 for Pb at the center.

However, as we shall presently see, in spite of the difference both
in the asymptotic behaviour and in the overall normalization,
starting from $y\sim 2$ the results of the evolution begin to
look practically identical for both choices of the initial function,
except for the overall scale $Q_s$ mentioned in the Introduction.

\section{The gluon density and dipole cross-sections}

We define the gluon density as in [7]:
\beq
\frac{\partial [xG(x,k^2,b)]}{\partial^2b\partial k^2}=
\frac{2N_c}{\pi g^2}k^2\nabla_k^2\phi(y,k,b)
\equiv\frac{2N_c}{\pi g^2}h(y,k,b),
\eeq
with $y=(N_c\alpha_s/\pi)\ln(1/x)$.
This definition follows the logic of the BFKL
approach in which it corresponds to the average of two gluon fields in the
nucleus, in the axial gauge adopted in this approach. It also naturally
appears in the expression for the structure function of the nucleus
(see [7]). There exist different definitions of the gluon density in the
literature (see [13]). We are not going to discuss the problem of a
"correct" gluon density here, since in any case it is not a directly
measurable quantity, but rather serves to calculate the latter
(see comments in [14]). From a pragmatic point of view
our definition is well supported, since all observables can be
directly related to (13).

The results of our evolution for the gluon density are shown in
Figures 1 and 2. Not to bind ourselves to a particular value of the coupling
constant $\alpha_s$, we rather present $h(y,k,b)$ as a function of
the rescaled rapidity $y$. The actual gluon density at physical rapidity is
obtained after rescaling both $h$ and $y$ according to Eqs. (2) and
(13).

In Fig. 1 we show the gluon densities at the first stage of the
evolution up to $y=1$, starting from the initial functions
corresponding to the Glauberized two-gluon exchange contribution (11)
(theoretical initial function, TIF) and to the
phenomenological cross-section
(12) (phenomenological initial function, PIF), respectively.
One observes a large difference between the two at the beginning of
the evolution, which, however, is gradually disappearing. Starting
from $y\sim 2$ the form of the gluon distribution becomes practically
identical for both initial functions. We introduce $Q_s(y,b)$ as the
the momentum at which the density reaches its maximum. For the two initial
functions a numerical fit gives Eq. (2)
with $a=-2.21$ $(-0.88)$ , $c=2.09$ $(2.04)$  and $d=0.69$ $(0.55)$
for TIF (PIF). Plotted against
$k/Q_s(y,b)$ the gluon distributions for both initial functions and
for all values of $b$ and $y>2$ practically fall onto a universal curve.
The degree of universality is illustrated in Fig. 2 where we present
the gluon densities for both TIF and PIF at $y=2.4$ and 4, and at
$b=0$ and $b=0.98 R_A$, for the Pb target.  Some
differences in the curves are certainly visible, which however diminish as $y$
grows.  The form of the universal curve can be
fitted by an expression
\beq
h(\xi)=
0.295\exp{\left(-\frac{(\xi-\xi_0)^2}{4\cdot 0.869}\right)},
\ \ \xi=\ln{k},\ \ \xi_0=\ln{Q_s(y,b)}
\eeq
(all momenta in GeV/c),
from which one observes that the density falls as a function of momentum
both at $k\to 0$ and $k\to \infty$.

The gluon density (13) is trivially related to the dipole cross-section on
the nucleus at fixed $b$:
\beq
\Phi(y,r,b)=1-\int\frac{d^2 k}{2\pi k^2}h(y,k,r)e^{i{\bf kr}}.
\eeq
So, to determine $\Phi$, all one has to do is to perform
a Bessel transform
of $h/k$.  Fig. 3 shows the dipole cross-sections at low rapidities.
At larger rapidities $y>2$ the scaling properties of $h$ imply that
$\Phi$ becomes a universal function of $rQ_s(y,b)$. Its form is
illustrated in Fig. 4, where we show $\Phi(r)$ for Pb at $b=0$ and
$y=3.4$ (for PIF).
It can be fitted with a formula
\beq
\Phi(r)=\left[1-\exp{\left(-\omega^2 r^2\right)^{1/\delta}}
\right]^\delta,\ \ \omega=198.5 \ {\rm GeV/c},
\ \ \delta=5.48.
\eeq

\section{The quark density}

The definition of the quark density can be taken from [13]:
\[
\frac{\partial [xq(x,l,b)]}{\partial^2l\partial^2b}
=\frac{\alpha_sQ^2}{(2\pi)^3}
\int_0^1d\alpha d^2b_1d^2b_2e^{-i{\bf l(b_1-b_2)}}\]\[
\left[\left(\alpha^2+(1-\alpha)^2\right)\epsilon^2\frac{{\bf b_1b_2}}{b_1b_2}
{\rm K}_1(\epsilon b_1){\rm K}_1(\epsilon b_2)+
4Q^2\alpha^2(1-\alpha)^2{\rm K}_0(\epsilon b_1){\rm K}_0(\epsilon b_2)
\right]\]\beq
\int\frac{d^2k}{(2\pi)^2}\frac{1}{k^2}
\frac{\partial [xG(x,k,b)]}{\partial^2b\partial k^2}
\left[1+e^{-i{\bf k(b_1-b_2)}}-e^{-i{\bf kb_1}}-e^{i{\bf kb_2}}\right],
\eeq
with $\epsilon^2=Q^2\alpha (1-\alpha)$.
This definition is based on the form of the interaction with the
target of a virtual current which splits into a $q\bar q$ pair.
For the small $x$ region one may also raise objections as to its
physical meaning, since then the interference diagram gives a
non-zero contribution [14]. However, for lack of a better definition,
we shall use Eq. (17).

Performing part of the integrations and using Eq. (13), we express
the quark density as
\beq
\frac{\partial [xq(x,l,b)]}{\partial^2l\partial^2b}=\pi^4f(y,l,b),\ \
\eeq
where $y=(N_c\alpha_s/\pi)\ln(1/x)$
and function $f$, independent of the coupling
constant, is given by an  integral over momenta:
\beq
f(y,l,b)=\frac{N_c}{8\pi^8}\int_0^{\infty}\frac{dk}{k}h(y,k,b)
I(l,k).
\eeq
Here $I$ is a sum of transversal ($T$) and longitudinal ($L$) parts,
with
\beq
I_T=Q^2\int_0^1
d\alpha\left[\alpha^2+(1-\alpha)^2\right]
\left[\left(2\epsilon^2+k^2\right)\chi\lambda-\epsilon^2\left(
\epsilon^2+l^2+k^2\right)\chi^3-
\epsilon^2\lambda^2\right]
\eeq
and
\beq
I_L=4Q^4\int_0^1d\alpha^2\alpha(1-\alpha)^2
\left[\left(\epsilon^2+l^2+k^2\right)\chi^3+\lambda^2-2\chi\lambda\right],
\eeq
where
\beq
\lambda=\frac{1}{\epsilon^2+l^2}\ ,\ \
\chi=\frac{1}{\sqrt{\left(\epsilon^2+(l+k)^2
\right)\left(\epsilon^2+(l-k)^2\right)}}\ .
\eeq

Note that the quark density defined by Eqs. (17)-(22) depends (weakly)
on the virtuality of the probe $Q^2$. Physically meaningful
results correspond to the limiting case $Q^2\to \infty$.
In this limit the longitudinal part of $I$ gives no contribution and
in the transversal part the integration over $\alpha$ is reduced to that
over
$\epsilon^2$:
\beq
Q^2\int_0^1d\alpha\left[\alpha^2+(1-\alpha)^2\right]\to
2\int_0^{\infty}d\epsilon^2,
\eeq
so that $I$ and $f$ become independent of $Q^2$.

These formulas give the quark density in terms of the gluon density,
that is, in terms of the function $h$, which we have found numerically.
Doing the integration  over $k$ in Eq. (19) also numerically, we obtain
function $f$ and therefore the quark density related to it by a factor
(see Eq. (18)). Our results for the quark density (in fact for $f$) are
presented in Figs. 5 and 6. As before, in Fig. 5 we show the quark
densities at the beginning of the evolution for the initial functions
TIF and PIF respectively. At higher $y>2$ and due to scaling properties
of the gluon density, the quark density also becomes a universal function
of $l/Q_s(y,b)$. Its form is illustrated in Fig. 6, where we show $f(l)$
for Pb at $b=0$ and $y=3.6$ (for PIF). It can
be fitted by
\beq
f(t)=\frac{N_c}{2 \pi^8} \frac{\left[1-\exp{\left(-t^{1/\beta}
\right)}
\right]^\beta}{t}\ ,\ \
t=2\left(\frac{l}{511.0\ {\rm GeV/c}}\right)^2,\ \ \beta=2.44.
\eeq
Note that at $l\to 0$ the quark densities tend to a constant value
in agreement with the prediction of A. H. Mueller [13]:
\beq
\frac{\partial [xq(x,l,b)]}{\partial^2l\partial^2b}\longrightarrow_{l\to 0}
\frac{N_c}{2\pi^4}\ ,
\eeq
which implies
\[ f(l)\longrightarrow_{l\to 0} 1.58\cdot 10^{-4};\]
at $l\to \infty$ they acquire a perturbative form ($\propto 1/l^2$).

\section{Discussion}

The new run of  numerical investigations of the BFKL fan diagram
equation for the gluon density in heavy nuclei fully confirms
our previous conclusions made in [7]. The key one is that at
rapidities of the order 10 the density forgets its initial form and
becomes a universal function  of momentum scaled by the
saturation momentum $Q_s(y,b)$. The latter grows as a power of energy
and as $A^{\alpha}$ with $\alpha$ of the order of 2/9.
Both the dipole-nucleus cross-section and the quark density behave in
agreement with a general idea of saturation, tending to a known
constant (unity for the dipole cross-section at fixed impact parameter)
as rapidity grows.

Numerical studies, as experimental ones, need confirmation from
independent groups. To our knowledge, up to now there appeared only two
papers devoted
to the solution of the non-linear BFKL equation, both of them for the nucleon
target. In [15] a simple Pad\'e
technique is used to solve the non-linear equation obtained
in [6]. Employing for the
linear case a solution which embodies both BFKL and DGLAP behaviour and
a running coupling constant, the authors present results for the
integrated gluon
distribution at Large Hadron Collider energies, finding a
suppression
effect, due to the non-linearity, of a factor $\sim 2$.
More appropriate for comparison with our results are those of [16], in
which an iteration
technique 
in coordinate space was used and results for the dipole cross-sections
were shown. The
chosen initial function is different from both our choices. It
attempts to include the DGLAP evolution at the initial stage.
We are not going to discuss here the viability of such an attempt
for the nucleus target. Still our results show that at rapidities
studied in [16], of the order of $12\div 15$, which with $\alpha_s=0.25$ mean
$y\sim 3.3\div 3.8$, the dipole cross-sections $\Phi(r)$ ($\tilde{N}(r)$ in the
notation of [16]) should become a universal function of $\bar{r}=rQ_s$.
The comparison indicates that although the general
behaviour of $\Phi$ found in [16] well agrees with ours, the universality in
the
$\bar{r}$-behaviour is not observed there. The slope of their curves in $\ln
\bar{r}$ is steeper than for our universal curve and this difference
grows with rapidity\footnote{After the first version of our work,
an
extension of the one in [16] to the nuclear case has appeared, [17].
As in [16],
the initial conditions are different from ours, but the authors find a
dependence $Q_s \propto A^{\alpha}$ with $\alpha=1/6$ at very small $x$,
quite similar to the one we obtain. Besides, they claim that the universality of
gluon distributions and dipole cross-sections is also observed in
their numerical solutions.}.

\section*{Acknowledgments}

M. A. B. is deeply thankful for attention and hospitality to the Departamento de
F\'{\i}sica of the Universidad de C\'ordoba,
where this work was done.
His work was also sponsored in part by grants of RFFI and
of NATO PST.CLG.976799. N. A. acknowledge financial support by CICYT of Spain
under contract
AEN99-0589-C02, and by Universidad de C\'ordoba.

\section*{References}
%

\noindent [1] A. H. Mueller, Nucl. Phys. {\bf B415} (1994) 373;
A. H. Mueller and B. Patel, Nucl. Phys. {\bf B425} (1994) 471.

\noindent [2] N. N. Nikolaev and B. G. Zakharov, Z. Phys. {\bf C64} (1994)
631.

\noindent [3] K. Golec-Biernat and M. W\"usthoff, Phys. Rev. {\bf D59} (1999)
014017; {\bf D60} (1999) 114023.

\noindent [4] L. V. Gribov, E. M. Levin and
M. G. Ryskin, Phys. Rept. {\bf
100} (1983) 1.

\noindent [5] I. I. Balitsky, Nucl. Phys. {\bf B463} (1996) 99;
hep-ph/9706411.

\noindent [6] Yu. V. Kovchegov, Phys. Rev. {\bf D60} (1999) 034008;
{\bf D61} (2000) 074018.

\noindent [7] M. A. Braun, Eur. Phys. J. {\bf C16} (2000) 337.

\noindent [8] L. McLerran and R. Venugopalan,
Phys. Rev. {\bf D49} (1994) 2233; 3352;
{\bf D50} (1994) 2225; {\bf D59} (1999) 094002; A. Ayala, J.
Jalilian-Marian, L. McLerran and R. Venugopalan, Phys. Rev.
{\bf D53} (1996) 458.

\noindent [9] J. Jalilian-Marian, A. Kovner, L. McLerran and H. Weigert,
Phys. Rev. {\bf D55} (1997) 5414;
J. Jalilian-Marian, A. Kovner,
A. Leonidov and
H. Weigert, Phys. Rev. {\bf D59} (1999) 014014; 034007
(Erratum: 099903); J. Jalilian-Marian, A. Kovner and H. Weigert, Phys. Rev.
{\bf D59} (1999) 014015.

\noindent [10] E. Iancu, A. Leonidov and L. McLerran, hep-ph/0011241.

\noindent [11] A. H. Mueller and J. Qiu, Nucl. Phys. {\bf B268} (1986) 427.

\noindent [12] E. M. Levin and K. Tuchin, Nucl. Phys. {\bf B573} (2000) 833;
hep-ph/0101275.

\noindent [13] A. H. Mueller, Nucl. Phys. {\bf B558} (1999) 285.

\noindent [14] M. A. Braun, hep-ph/0010041.

\noindent [15] M. A. Kimber, J. Kwieci\'nski and A. D. Martin,
hep-ph/0101099.

\noindent [16] M. Lublinsky, E. Gotsman,
E. M. Levin and U. Maor, hep-ph/0102321.

\noindent [17] E. M. Levin and M. Lublinsky,
hep-ph/0104108.

\section*{Figure captions}
%

\noindent{\bf Fig. 1:}
The gluon densities at the first stage of the evolution $y<1$,
for Pb target at $b=0$.
Solid and dashed curves show the densities evolved from TIF
and PIF at $y=0$ respectively. Curves from left to right correspond to
$y=0.0$, 0.4 and 1.0.

\noindent{\bf Fig. 2:}
Gluon densities  at $y=2.2$ and $4.0$, for Pb target at $b=0$ and
$b=0.98 R_A$, both for TIF (solid curves) and PIF (dashed curves)
as initial functions, plotted
against $k/Q_s(y,b)$.

\noindent{\bf Fig. 3:} The dipole cross-sections $\Phi(r)$
at the first stage of the evolution $y<1$,
for Pb target at $b=0$.
Solid and dashed curves show the cross-sections for TIF
and PIF as initial functions respectively. Curves from right to left
correspond to $y=0.0$, 0.4 and 1.0.

\noindent{\bf Fig. 4:} The dipole cross-sections $\Phi(r)$ at $y=3.4$ for
Pb at $b=0$, with PIF as initial function.

\noindent{\bf Fig. 5:}
The quark densities  $f(l)$
at the first stage of the evolution $y<1$,
for Pb target at $b=0$.
Solid and dashed curves show the quark densities for TIF
and PIF as initial functions respectively. Curves from left to right
correspond to $y=0.0$, 0.4 and 1.0.

\noindent{\bf Fig. 6:}
The quark densities $f(l)$ at $y=3.6$ for
Pb at $b=0$, with PIF as initial function.

\newpage
\centerline{\bf \large Figures:}

\begin{figure}[htb]
\begin{center}
\epsfig{file=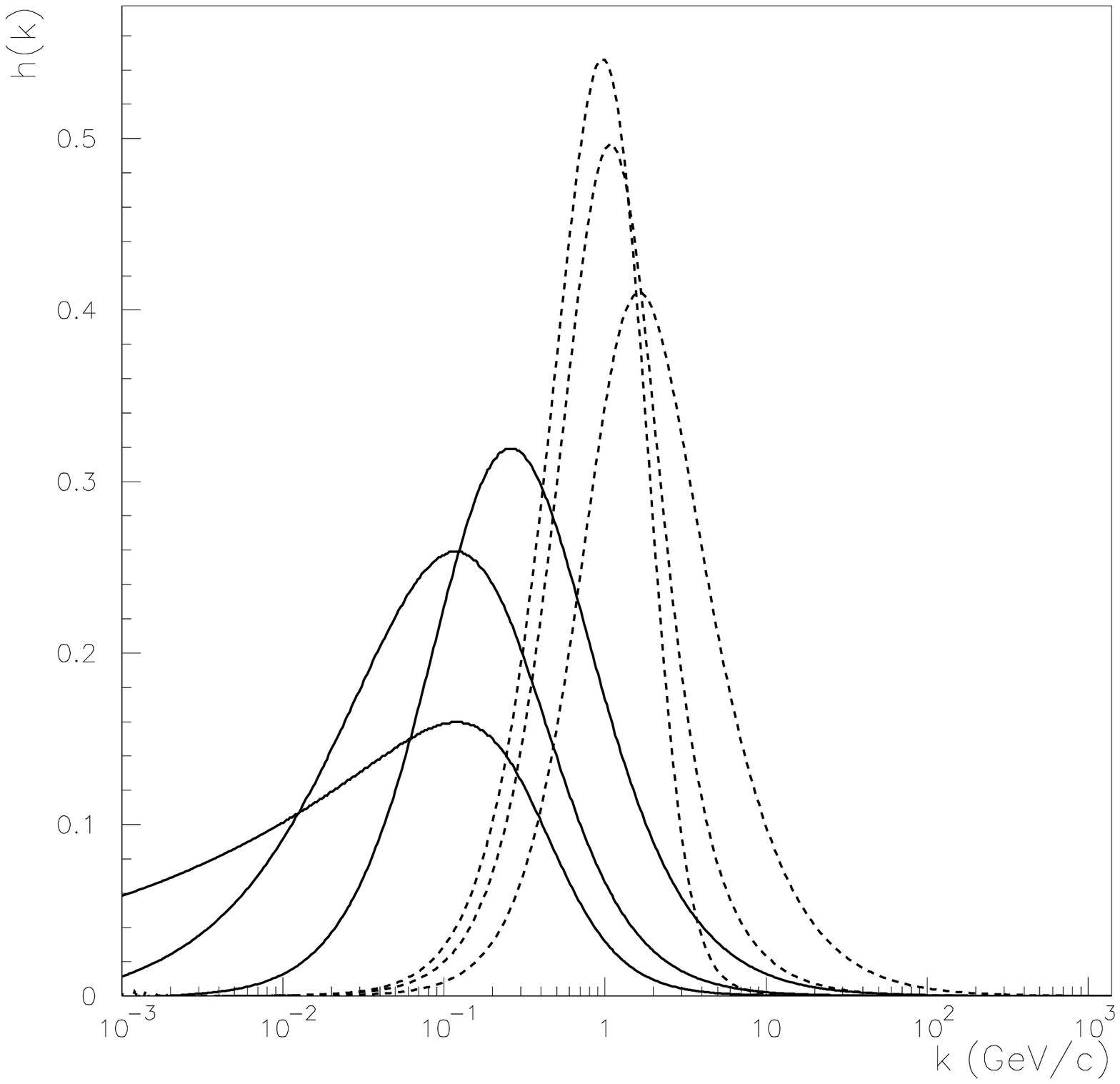,width=15.5cm}
\end{center}
\vskip -1.0cm
\caption{}
\label{fig1}
\end{figure}

\newpage

\begin{figure}[htb]
\begin{center}
\epsfig{file=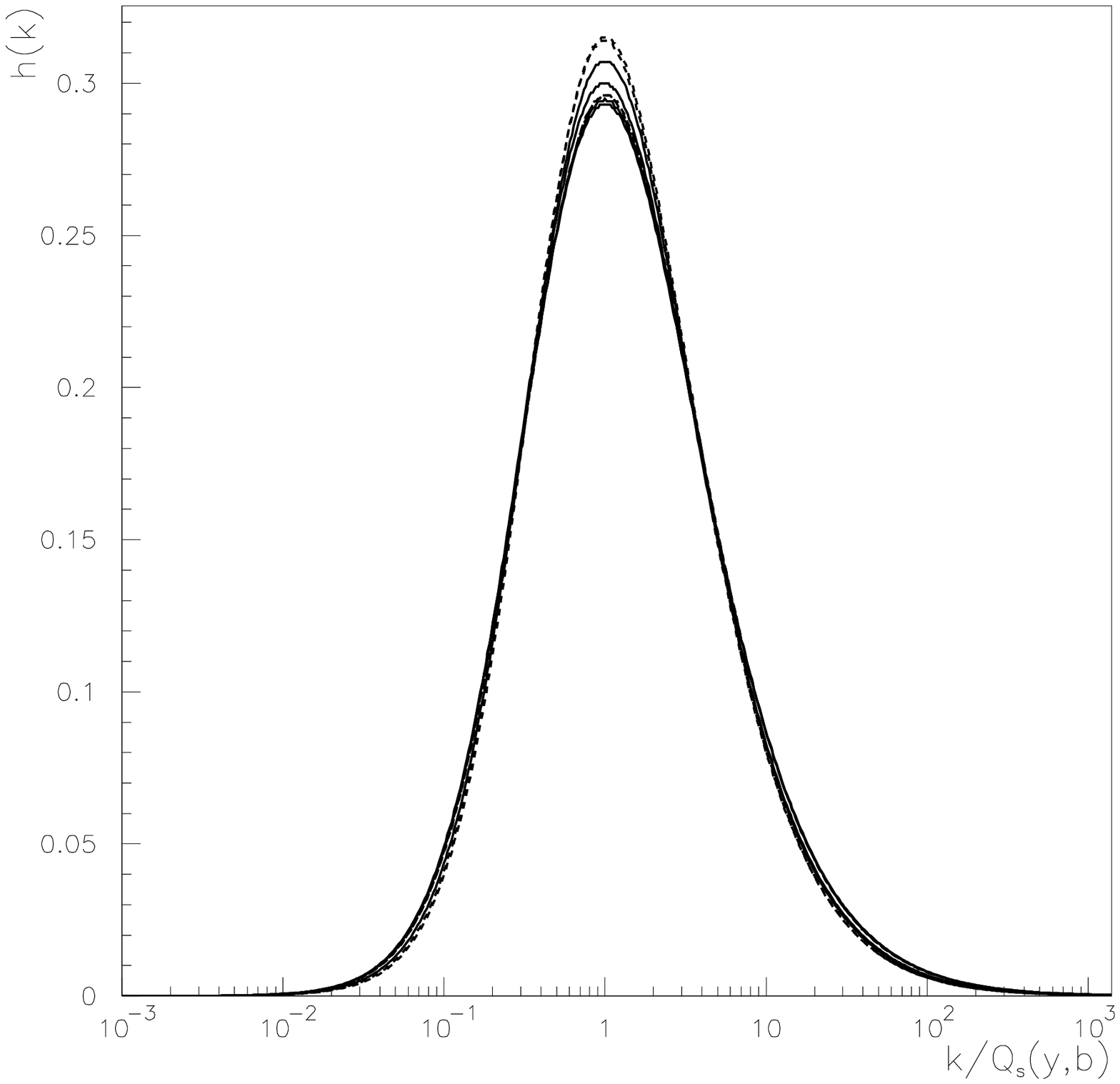,width=15.5cm}
\end{center}
\vskip -1.0cm
\caption{}
\label{fig2}
\end{figure}

\newpage

\begin{figure}[htb]
\begin{center}
\epsfig{file=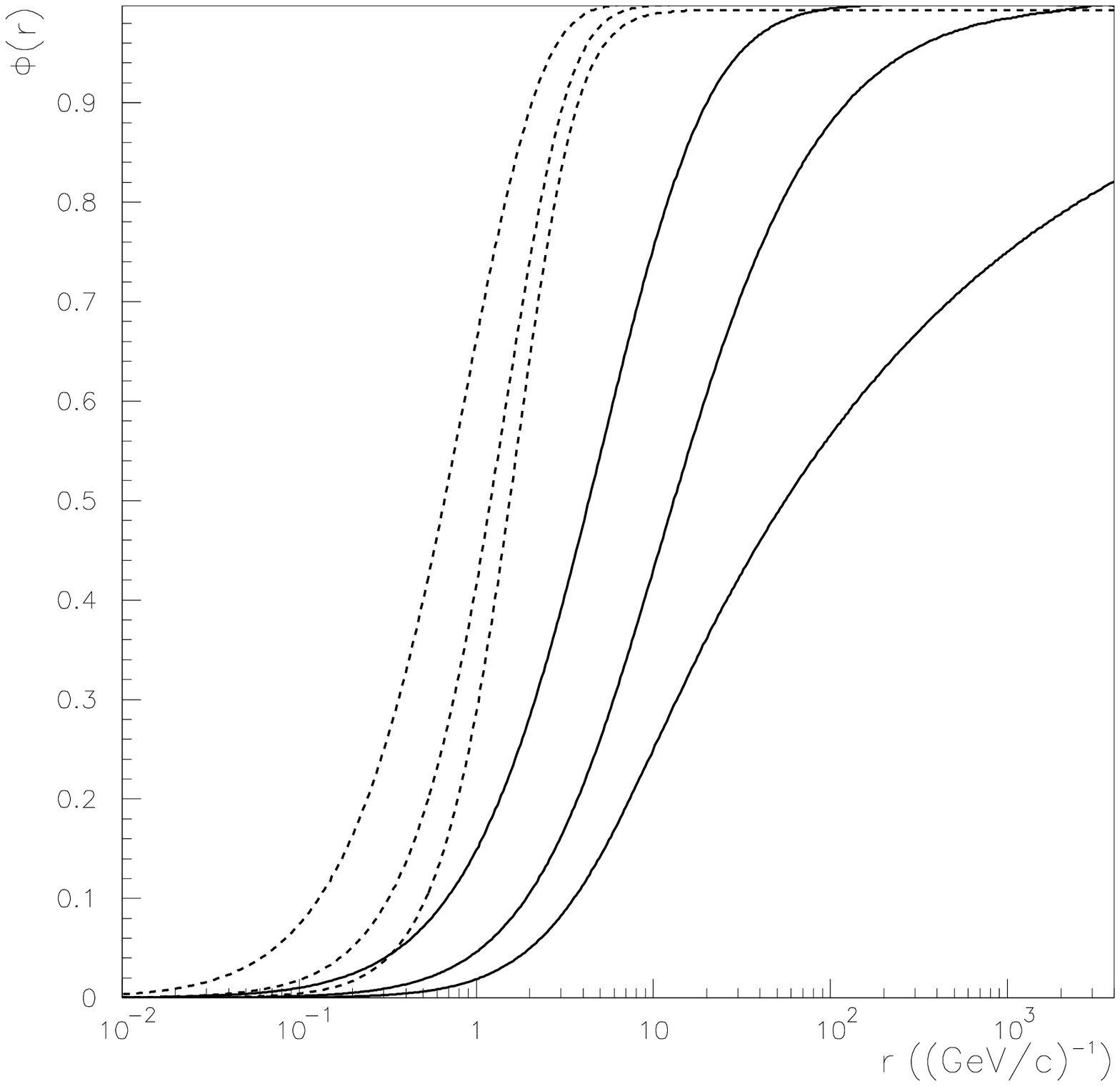,width=15.5cm}
\end{center}
\vskip -1.0cm
\caption{}
\label{fig3}
\end{figure}

\newpage

\begin{figure}[htb]
\begin{center}
\epsfig{file=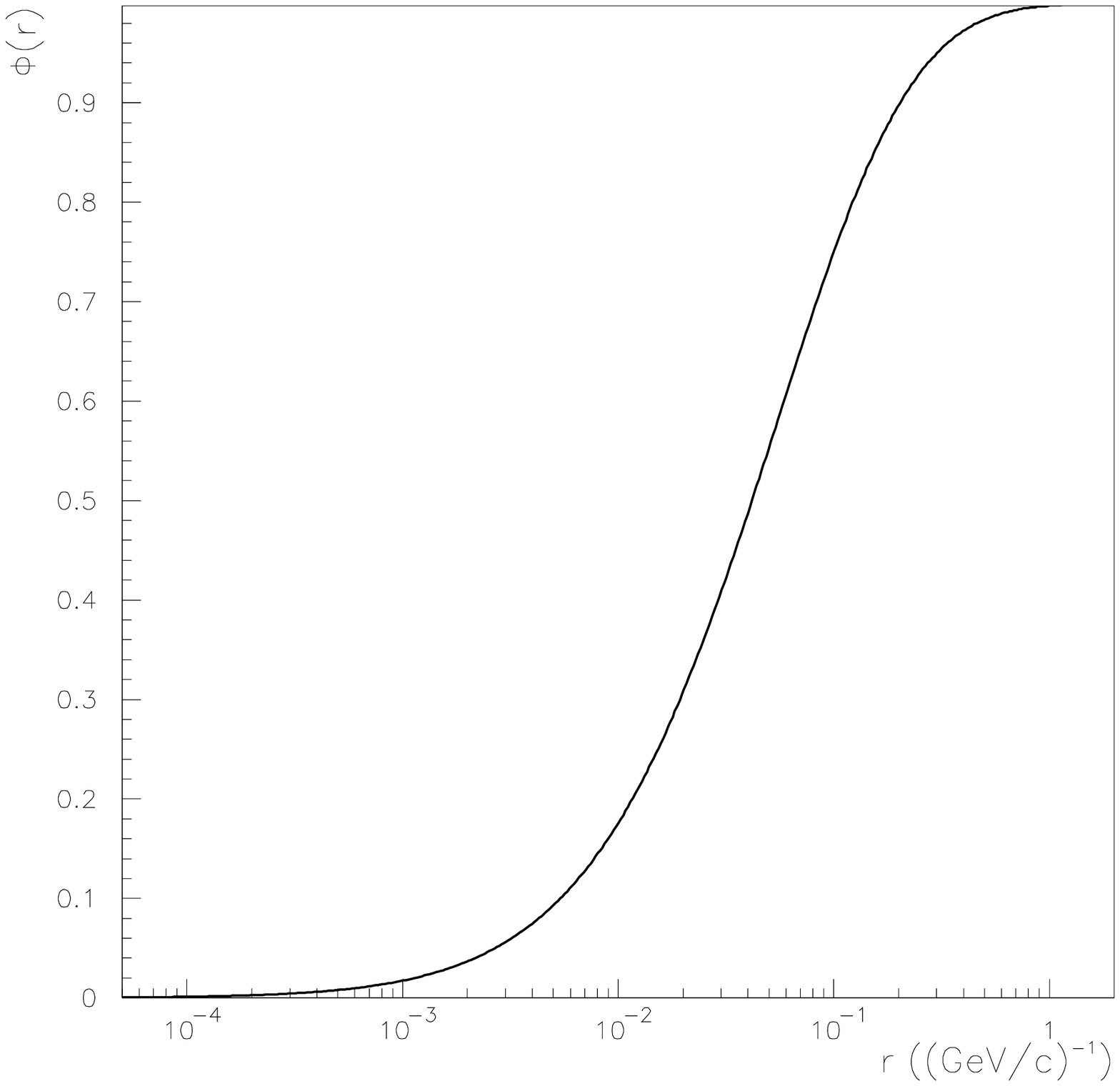,width=15.5cm}
\end{center}
\vskip -1.0cm
\caption{}
\label{fig4}
\end{figure}

\newpage

\begin{figure}[htb]
\begin{center}
\epsfig{file=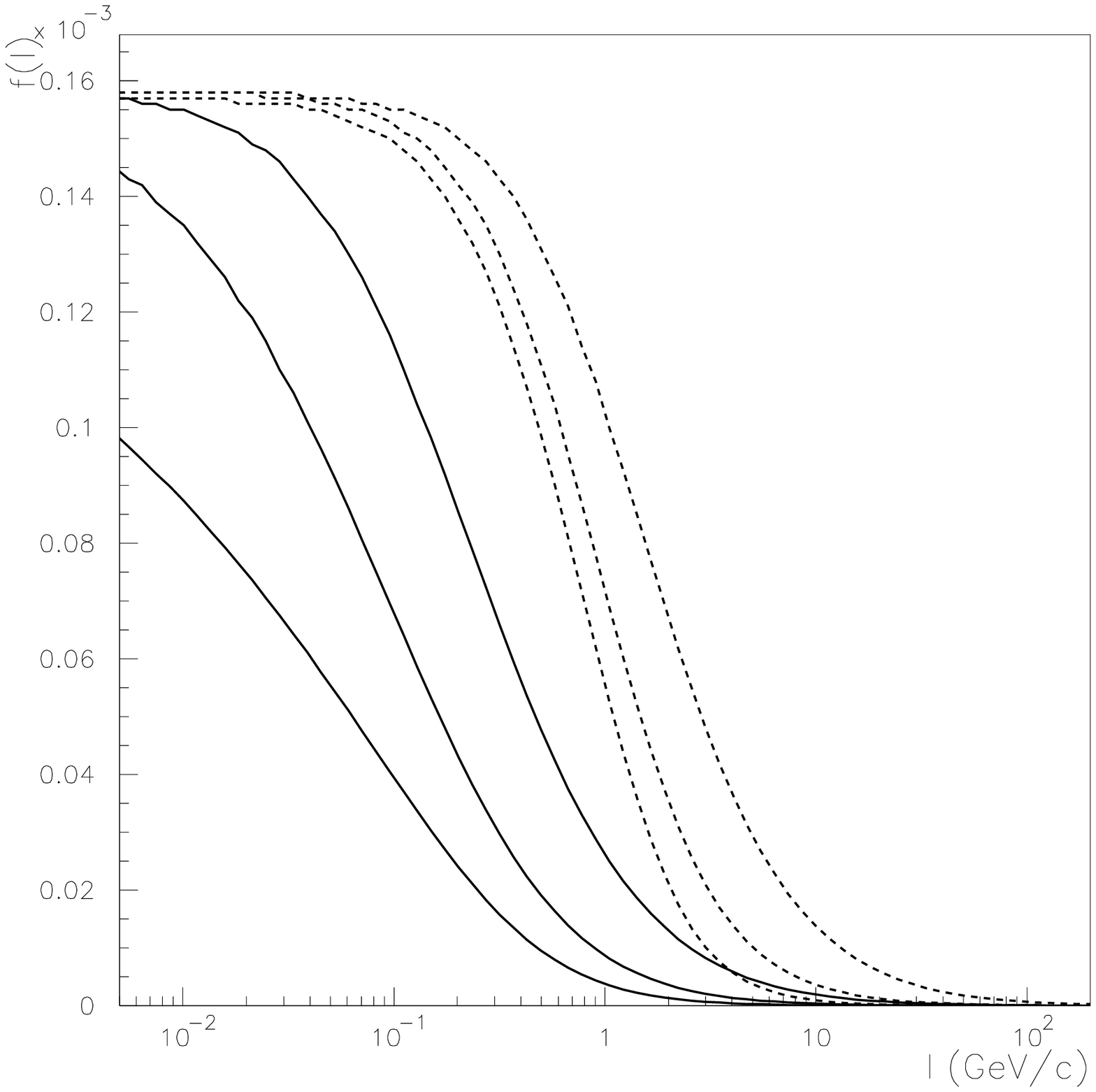,width=15.5cm}
\end{center}
\vskip -1.0cm
\caption{}
\label{fig5}
\end{figure}

\newpage

\begin{figure}[htb]
\begin{center}
\epsfig{file=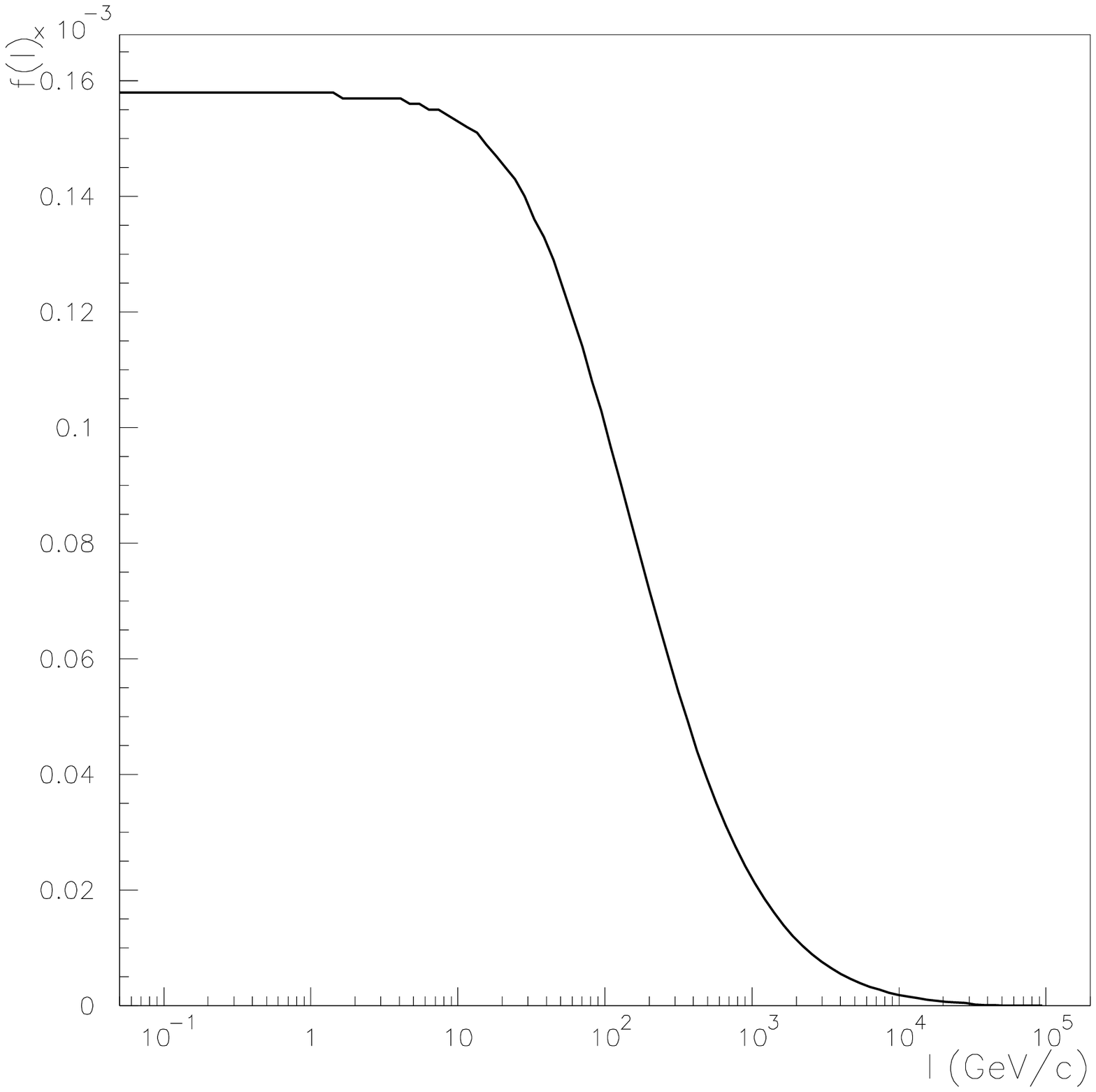,width=15.5cm}
\end{center}
\vskip -1.0cm
\caption{}
\label{fig6}
\end{figure}

\end{document}